\documentclass[epj]{svjour}

\usepackage{fancyhdr}
\usepackage{amsmath,amssymb}
\def\makeheadbox{{%  remove EPJC header
 }}
\def\mytitle{My title} 
\def\myauthors{My name}  
\def\mytype{My type of session}
\def\mysession{My session}
\def\mytitle{Higher Derivative Corrections in M-theory} %Put your title here!
\def\myauthors{Yoshifumi Hyakutake}    %Put your name here!
\def\mytype{Contributed Talk}    
\def\mysession{Theoretical Models}

\newcommand{\ba}{\begin{alignat}{3}}

\newcommand{\dl}{\delta}
\newcommand{\ep}{\epsilon}

\newcommand{\tf}{\tfrac}
\newcommand{\mc}{\mathcal}

\begin{document}
\title{Higher Derivative Corrections in M-theory via Local Supersymmetry}
%\subtitle{Do you have a subtitle?\\ If so, write it here}
\author{Yoshifumi Hyakutake%\inst{1}
% \thanks is optional - remove next line if not needed
\thanks{\emph{Email:} hyaku@het.phys.sci.osaka-u.ac.jp}%
% \and
% Second author\inst{2}% etc
% \thanks is optional - remove next line if not needed
%\thanks{\emph{Present address:} Insert the address here if needed}%
}                     % Do not remove
%
%\offprints{}          % Insert a name or remove this line
%
\institute{Institute for Higher Education Research and Practice, Osaka University, \\
Machikaneyama 1-16, Toyonaka, Osaka 560-0043, Japan
%\and the second institute
}
%
%\date{Received: date / Revised version: date}
% The correct dates will be entered by Springer
\date{}
\abstract{
We investigate higher derivative corrections in M-theory by applying
Noether's method. Cancellation of variations, which contain linear terms in 4-form field
strength, under local supersymmetry is executed with the aid of computer programming. 
Structure of $R^4$ terms is uniquely determined and exactly matches with one-loop effective terms
in type IIA superstring theory.
 % end of PACS codes
} %end of abstract
\maketitle
%DO NOT REMOVE THIS LINE
%

\section{Introduction}
\label{intro}

Superstring theory is promising as a theory of quantum gravity, and
interactions of strings are described by supergravity in the low energy limit\cite{Y,SS}.
Scattering amplitudes of strings also contain higher derivative interactions which give 
stringy or quantum corrections to supergravity\cite{SS}. 

Among the higher derivative terms, quartic terms of Riemann tensor, $R^4$ terms, in type II superstring 
theories are considerably investigated in the last twenty years from various viewpoints.
Especially tree level effective action which include $R^4$ terms is obtained by
combining results of scattering amplitude of four gravitons and 4-loop 
computation in non-linear sigma model\cite{GW,GVZ}. Effective action at one-loop order also
contains the $R^4$ terms. We will review some aspects of $R^4$ terms in sect.~\ref{sec:IIA}.

By lifting the effective action in type IIA superstring theory to 11 dimensions,
higher derivative corrections in M-theory are revealed. 
It is also possible to directly confirm the existence of higher derivative terms
in eleven dimensions from the scattering amplitudes of 
superparticles\cite{GV,GGV,RT,AGV} or superspace formalism\cite{CGNN,CGNN2,HSS,HT,Ra}.

Without relying on scattering amplitudes, higher derivative corrections in superstring 
theories are determined by imposing local supersymmetry\cite{RSW1,RSW2,Su,PVW,HO1,HO2,H1}.
This paper is a summary of works of refs. \cite{HO1,HO2,H1}.
The structure of higher derivative terms in M-theory is investigated by applying Noether's method. 
Variations under local supersymmetry which are linearly dependent on 4-form field strength
are cancelled with the aid of a computer programming\footnote{I employed mathematica code, 
GAMMA.m, built by U. Gran\cite{Gr}.}. 
We will see that the local supersymmetry is powerful enough to 
determine the structure of $R^4$ terms uniquely. A part of $R^3F^2$ terms is also determined.

The contents of our paper is as follows. In section \ref{sec:IIA}, we briefly summarize
higher derivative corrections in type IIA superstring theory.
In section \ref{sec:M}, higher derivative corrections in M-theory are investigated via 
local supersymmetry. We find that the structure of $R^4$ terms is uniquely determined 
and is consistent with one-loop scattering amplitudes in type IIA superstring theory. 
Section \ref{sec:Discuss} is devoted to conclusions and discussions.

\section{Higher Derivative Corrections in Type IIA Superstring Theory}
\label{sec:IIA}

It is well known that the low energy limit of superstring theory
is described by supergravity. This can be shown explicitly by
calculating the scattering amplitudes of massless states.
For example, low energy limit of amplitude of three gravitons 
at tree level correctly reproduces three points vertex of gravitons
in the supergravity Lagrangian.

Scattering amplitudes of massless states in superstring theory, however, include corrections 
to the supergravity Lagrangian. These terms contain positive power of string length $\ell_s$ 
and string coupling constant $g_s$, and are called higher derivative corrections.
In type IIA superstring theory, one of those corrections appear in amplitude of 
four gravitons which is written by
\ba
  \mc{M}_4^{\text{tree}} &= - \frac{1}{2\kappa_{10}^2} \frac{\ell_s^6}{2^9 \cdot 4!} \, 
  T(s,t,u) \, t_8 t_8 R^4.
\end{alignat}
Here $2\kappa_{10}^2 = (2\pi)^7 \ell_s^8 g_s^2$, $t_8$ is composed of 4 Kronecker deltas, 
and $t_8 t_8 R^4$ is an abbreviation of a product of two $t_8$ tensors and four Riemann tensors.
Explicit expression can be found in ref.~\cite{HO2}.
$T(s,t,u)$ is a function of Mandelstam variables and defined by using gamma functions as
\ba
  T(s,t,u) &= \frac{\Gamma(-\frac{\ell_s^2}{4}s)\Gamma(-\frac{\ell_s^2}{4}t)\Gamma(-\frac{\ell_s^2}{4}u)}
  {\Gamma(1+\frac{\ell_s^2}{4}s)\Gamma(1+\frac{\ell_s^2}{4}t)\Gamma(1+\frac{\ell_s^2}{4}u)} \notag
  \\
  &\sim \frac{2^6}{\ell_s^6 stu} + 2 \zeta(3) + \cdots.
\end{alignat}
The second line in the above equation gives the low energy limit of the amplitude.
The first term in that line just corresponds to amplitudes of exchanging four gravitons in $s$, $t$ and
$u$ channels in type IIA supergravity\cite{Sa}. The second term contain a Riemann zeta function $\zeta(3)$,
which is irrational, and contributes as a stringy correction to the supergravity.
In this way it is basically possible to derive higher derivative corrections from string scattering
amplitudes, and terms which involve fourth power of Riemann tensors become
\ba
  \mc{L}_{\text{tree}} &= - \frac{e^{-2\phi}}{2\kappa_{10}^2} \frac{\ell_s^6}{2^8 \cdot 4!} 
  \zeta(3) \notag
  \\
  &\quad\,
  \times \big(t_8 t_8 R^4 + \tf{1}{4\cdot 2!}\ep_{10}\ep_{10}R^4 \big). \label{eq:tree}
\end{alignat}
The second term is first derived by calculations in non-linear sigma model\cite{GVZ}.
By counting mass dimensions, we see that above Lagrangian should be order of $\ell_s^6$.

At 1-loop level, 0, 1, 2, 3 points amplitudes of massless states are zero, and 
higher derivative corrections first arise out of amplitude of four gravitons.
\ba
  \mc{M}_4^{\text{1-loop}} &= \frac{1}{2\kappa_{10}^2} \frac{\pi g_s^2 \ell_s^6}{2^8 \cdot 4!} \, 
  I(s,t,u) \, t_8 t_8 R^4.
\end{alignat}
$I(s,t,u)$ is an integral of a modular function on a torus and approximated in the low energy limit as
\ba
  I(s,t,u) \sim \frac{\pi}{3}.
\end{alignat}
Finally the effective action which contain fourth power of Riemann tensor at 1-loop level become
\ba
  \mc{L}_{\text{1-loop}} &= \frac{1}{2\kappa_{10}^2} 
  \frac{\ell_s^6}{2^8 \cdot 4!} \frac{\pi^2}{3}g_s^2 \label{eq:loop}
  \\
  &\quad\,
  \times \big(t_8 t_8 R^4 - \tf{1}{4\cdot 2!}\ep_{10}\ep_{10}R^4
  - \tf{1}{6} \ep_{10} t_8 BR^4 \big). \notag
\end{alignat}
The last term comes from 5 points amplitude of antisymmetric tensor field and 4 gravitons.
Note that the signs in front of $\ep_{10}\ep_{10}R^4$ are opposite.
Those are sensitive to the chirality of the theory.

So far we have concentrated on terms which involve fourth power of the Riemann tensor.
Compared with gravitons, our knowledge on higher derivative terms which consist of NS $B$ 
field and R-R potentials is poor. In order to determine these terms, it is necessary to 
calculate more than 5 points amplitudes and extract their low energy limit correctly.
This is very tough work. In the next section we consider higher derivative corrections 
in M-theory and try to determine the structure of these terms only by employing local supersymmetry.

\section{Higher Derivative Corrections in M-theory}
\label{sec:M}

The string coupling constant in type IIA superstring theory is identified with 11th circular
direction with a radius $R=g_s\ell_s$, and strong coupling limit of this theory is described
by 11 dimensional M-theory. The low energy limit of M-theory is described by 11 dimensional supergravity,
and gravity multiplet consists of a vielbein $e^a{}_\mu$, a 3-form field $A$ and 
a Majorana gravitino $\psi_\mu$. In M-theory a supermembrane plays a fundamental role and often
capture nonperturbative aspects of superstring. However, quantization of supermenbrane is incomplete, 
so it is impossible to obtain higher derivative terms by evaluating scattering amplitudes 
of supermembranes.

Therefore we should employ another method to determine the structure of higher derivative corrections
in M-theory. Since the gravity multiplet is simple, it will be possible to fix the structure
by local supersymmetry. In this section we will apply Noether's method with the aid of computer
programming, and show that terms which involve fourth power of the Riemann tensor, which
are abbreviated as $[eR^4]$, can be determined uniquely.

First let us classify independent terms in $[eR^4]$. Here we only consider terms 
which do not contain Ricci tensors or scalar curvatures. Because of this restriction,
four indices of one Riemann tensor should be contracted with those of other Riemann tensors.
Then it is convenient to consider a $4 \times 4$ symmetric matrix whose $(i,j)$ component represent 
a number of contracted indices between $i$-th and $j$-th Riemann tensors. 
The sum of each column should be 4, which is the number of indices in the Riemann tensor, and 
the symmetric matrix is controlled by two parameters as
\ba
  \begin{pmatrix}
    0     & a     & b     & 4-a-b \\
    a     & 0     & 4-a-b & b     \\
    b     & 4-a-b & 0     & a     \\
    4-a-b & b     & a     & 0
  \end{pmatrix},
\end{alignat}
where $0 \leq a,b,a+b \leq 4$.
Without loss of generality we set $4-a-b \leq b \leq a$, since this is always possible by 
properly exchanging positions of Riemann tensors. Then it is easy to see that there
are 4 possible matrices.
\ba
  \begin{pmatrix}
    0 & 4 & 0 & 0 \\
    4 & 0 & 0 & 0 \\
    0 & 0 & 0 & 4 \\
    0 & 0 & 4 & 0
  \end{pmatrix},
    \begin{pmatrix}
    0 & 3 & 1 & 0 \\
    3 & 0 & 0 & 1 \\
    1 & 0 & 0 & 3 \\
    0 & 1 & 3 & 0
  \end{pmatrix},  \begin{pmatrix}
    0 & 2 & 2 & 0 \\
    2 & 0 & 0 & 2 \\
    2 & 0 & 0 & 2 \\
    0 & 2 & 2 & 0
  \end{pmatrix},
    \begin{pmatrix}
    0 & 2 & 1 & 1 \\
    2 & 0 & 1 & 1 \\
    1 & 1 & 0 & 2 \\
    1 & 1 & 2 & 0
  \end{pmatrix},
\end{alignat}
Patterns of contractions of two Riemann tensors are considerably restricted by the 
properties of the Riemann tensor, and possible expressions are given by
\ba
  &R_{abcd}R_{abcd}, \quad R_{abcx}R_{abcx}, \notag
  \\
  &R_{abxx}R_{abxx}, \quad R_{axbx}R_{axbx},
\end{alignat}
where $x$ represents a blank and $a,b,c$ and $d$ are contracted indices.
By using the properties of the Riemann tensor we see that there are 7 independent terms in $[eR^4]$.
\ba
  &R_{abcd}R_{abcd}R_{efgh}R_{efgh}, \; R_{abcg}R_{abch}R_{defg}R_{defh}, \notag
  \\
  &R_{abcd}R_{abef}R_{cdgh}R_{efgh}, \; R_{acbd}R_{aebf}R_{cgdh}R_{egfh}, \notag
  \\
  &R_{abef}R_{abgh}R_{cedg}R_{cfdh}, \; R_{abef}R_{abgh}R_{cedg}R_{chdf}, 
  \\
  &R_{aebf}R_{agbh}R_{cedg}R_{cfdh}. \notag
\end{alignat}
This procedure is systematic and it is possible to execute this by employing
computer programming. In order to supersymmetrize terms in $[eR^4]$, it is necessary
to add more possible terms in the effective action, and the result by computer programming 
is summarized as
\ba
  &B_1=[eR^4]_7, &\quad
  &B_{11}=[e\ep_{11}AR^4]_2, \notag
  \\
  &F_1=[eR^3\bar{\psi}\psi_2]_{92}, &\quad
  &F_2=[eR^2\bar{\psi}_2D\psi_2]_{25}.
\end{alignat}
Here $\ep_{11}$ is totally antisymmetric tensor in 11 dimensions, and
$\psi_2$ represents a field strength of the Majorana gravitino.
Each number in the subscript shows that of independent terms.
Details of these expressions are found in ref. \cite{HO2}.

Supersymmetric transformations of massless fields are the same as those
of 11 dimensional supergravity, and are abbreviated as\cite{CJS}
\ba
  \dl e = [\bar{\ep}\psi], \quad \dl\psi = [D\ep], \quad \dl A = [\bar{\ep}\psi].
\end{alignat}
$\ep$ is a parameter which belongs to a Majorana spinor representation.
Then variations of $B_1$, $B_2$, $F_1$ and $F_2$ are expanded by terms in 3 classes,
$V_1$, $V_2$ and $V_3$.
\ba
  V_1 &= [eR^4 \bar{\ep}\psi]_{116}, \notag
  \\
  V_2 &= [eR^2DR\bar{\ep}\psi_2]_{88}, 
  \\
  V_3 &= [eR^3\bar{\ep}D\psi_2]_{40}. \notag
\end{alignat}
Then cancellation mechanism of these variations are summarized in the following table.
\ba
  \dl B_1 &= V_1 \oplus V_2, \notag
  \\
  \dl B_{11} &= V_1, \notag
  \\
  \dl F_1 &= V_1 \oplus V_2 \oplus V_3, \label{eq:tb1}
  \\
  \dl F_2 &= \qquad\; V_2 \oplus V_3. \notag
\end{alignat}
Local supersymmetry requires that the right hand side in the above table should be zero.
Thus we obtain 244 equations among 126 variables.
After miraculous cancellation, the bosonic part is determined as
\ba
  \mc{L} &= a \big( t_8 t_8 R^4 + \tf{1}{4!}\ep_{11}\ep_{11}R^4 \big) \notag
  \\
  &\quad\,
  + b\big( t_8 t_8 R^4 - \tf{1}{4!}\ep_{11}\ep_{11}R^4 - \tf{1}{6} \ep_{11} t_8 AR^4 \big),
\end{alignat}
where $a$ and $b$ are free parameters.
The first line matches with the result of tree level scattering amplitude (\ref{eq:tree}),
and the second line is consistent with 1-loop amplitude (\ref{eq:loop}) in type IIA
superstring theory.

So far we have neglected terms which involve 4-form field strength.
Thus the next step is to examine the invariance under local supersymmetry up to $\mc{O}(F^2)$.
In order to execute the cancellation to this order, we have to add
\ba
  B_{21} &= [eR^3F^2]_{30}, &\; B_{22} &= [eR^2 (DF)^2]_{24}, \notag
  \\
  F_{11} &= [eR^3F\bar{\psi}\psi]_{447}, &\; F_{12} &= [eR^2F\bar{\psi}_2\psi_2]_{190}, 
  \\
  F_{13} &= [eR^2DF\bar{\psi}\psi_2]_{614}, &\; F_{14} &= [eRDF\bar{\psi}_2D\psi_2]_{113}. \notag
\end{alignat}
Variations of these terms are expanded by terms in 6 classes.
\ba
  V_{11} &= [eR^2DRF\bar{\ep}\psi]_{1563}, &\; V_{12} &= [eR^3F\bar{\ep}\psi_2]_{513}, 
  \\
  V_{13} &= [eR^3DF\bar{\ep}\psi]_{995}, &\; V_{14} &= [eRDRDF\bar{\ep}\psi_2]_{371}, \notag
  \\
  V_{15} &= [eR^2DF\bar{\ep}D\psi_2]_{332}, &\; V_{16} &= [eR^2DDF\bar{\ep}\psi_2]_{151}. \notag
\end{alignat}
Cancellation mechanism of variations which are linear in $F$ is summarized in the following table.
\ba
  \dl B_1 &= V_{11}, \notag
  \\
  \dl B_{11} &= V_{11} \qquad\;\;\, \oplus V_{13}, \notag
  \\
  \dl F_1 &= \qquad\;\;\, V_{12} \oplus V_{13}, \notag
  \\
  \dl F_2 &= \qquad\;\;\, V_{12} \qquad\qquad V_{14} \oplus V_{15}, \notag
  \\
  \dl B_{21} &= V_{11} \qquad\;\; \oplus V_{13}, \label{eq:tb2}
  \\
  \dl B_{22} &= \qquad\qquad\qquad\qquad V_{14} \qquad\;\; \oplus V_{16}, \notag
  \\
  \dl F_{11} &= V_{11} \oplus V_{12} \oplus V_{13}, \notag
  \\
  \dl F_{12} &= \qquad\;\;\, V_{12}, \notag
  \\
  \dl F_{13} &= \qquad\qquad\quad\; V_{13} \oplus V_{14} \oplus V_{15} \oplus V_{16}, \notag
  \\
  \dl F_{14} &= \qquad\qquad\qquad\qquad V_{14} \oplus V_{15} \oplus V_{16}. \notag
\end{alignat}
Requirement of local supersymmetry in (\ref{eq:tb1}) and (\ref{eq:tb2}) 
gives 4169 equations among 1544 variables. From this cancellation we obtain only one solution,
\ba
  \mc{L} &= b\big( t_8 t_8 R^4 - \tf{1}{4!}\ep_{11}\ep_{11}R^4 - \tf{1}{6} \ep_{11} t_8 AR^4 \big) \notag
  \\
  &\quad\,
  + ([R^3F^2] \oplus [R^2(DF)^2]). \label{eq:result}
\end{alignat}
Therefore by imposing local supersymmetry, the structure of $[eR^4]$ is uniquely determined.
The second line in the above equation contain 2 parameters and explicit forms will be reported 
elsewhere\cite{H1}.
The coefficient $b$ is fixed by comparing with eq.~(\ref{eq:loop}).
\ba
  b = \frac{1}{2\kappa_{11}^2} \frac{\ell_p^6}{2^8 \cdot 4!} \frac{\pi^2}{3},
\end{alignat}
where $2 \kappa_{11}^2 = (2\pi)^8\ell_p^9$ and $\ell_p = g_s^{1/3} \ell_s$.

\section{Conclusions and Discussions}\label{sec:Discuss}

The higher derivative terms in M-theory are investigated by applying
Noether's method. The cancellation of the variations under local supersymmetry
is examined to the order linear in the 4-form field strength $F$. 
Since the calculations are hard, we heavily employed the computer
programming to check the cancellation.

The bosonic part of the ansatz consists of 63 terms, $B_{1}$, $B_{11}$, $B_{21}$ and $B_{22}$, and 
the fermionic part of the ansatz does of 1481 terms, $F_{1}$, $F_{2}$, $F_{11}$,
$F_{12}$, $F_{13}$ and $F_{14}$. The variations of the ansatz are expanded by 4169 terms,
$V_{1}$, $V_{2}$, $V_{3}$, $V_{11}$, $V_{12}$, $V_{13}$, $V_{14}$, $V_{15}$ and $V_{16}$.
By requiring the cancellation of these variations, we obtain 4169 linear equations among
1544 coefficients in the ansatz.
Then the coefficients of the bosonic part is solved as eq.~(\ref{eq:result}).
Remarkably the structure of the $R^4$ terms is uniquely determined by the requirement 
of the local supersymmetry. This result exactly matches with the fact that the one-loop effective 
action in the type IIA superstring theory survives after taking $g_s \to \infty$.
When $g_s$ is finite, 11th direction is compactified on the circle.
Tree level $R^4$ terms in type IIA superstring theory arise after summing over
non-zero modes of Kaluza-Klein mass spectrum\cite{GGV}. From these arguments, we see
that there is no term which corresponds to amplitudes at 2-loop or more in type IIA superstring theory.
This gives proof of vanishing theorem via local supersymmetry.

The result by computer programming shows that $R^3F^2$ terms are governed by two parameters.
It is interesting to compare the result with that obtained by the scattering amplitudes in 
type IIA superstring theory\cite{PVW2}.
It is also important to check the consistency to the result obtained by 
the other methods\cite{Ra,DS,KMS}.

As a conclusion the local supersymmetry seems to determine the structure of the higher derivative
corrections in M-theory uniquely. Similar statement can be found in the context of 
D-particle dynamics\cite{KM}. We will succeed the procedure executed in this paper and determine 
the structure of the higher derivative corrections in M-theory completely. 
After the determination of the action, corrections to black $p$-branes
or cosmology will become interesting future directions\cite{FO,My,MO}.

\section*{Acknowledgement}

I would like to thank Jose Figueroa-O'Farrill for interesting discussions.
This research was partially supported by the Ministry of Education, Science, 
Sports and Culture, Grant-in-Aid for Young Scientists (B), 19740141, 2007, and 
by The 21st Century COE Program "Towards a New Basic Science; Depth and Synthesis.''

%
% BibTeX users please use
% \bibliographystyle{}
% \bibliography{}
%
% Non-BibTeX users please use

\end{document}